\documentclass[lettersize,journal]{IEEEtran}
\usepackage{amsmath,amsfonts}
\usepackage{algorithmic}
\usepackage{array}
\usepackage{textcomp}
\usepackage{multirow}
\usepackage{color}
\usepackage{stfloats}
\usepackage{url}
\usepackage{verbatim}
\usepackage{subfigure} 
\usepackage{graphicx}
\usepackage{booktabs}
\usepackage{svg}
\usepackage{bm}
\def\BibTeX{{\rm B\kern-.05em{\sc i\kern-.025em b}\kern-.08em
    T\kern-.1667em\lower.7ex\hbox{E}\kern-.125emX}}
\usepackage{balance}
\usepackage[numbers,sort&compress]{natbib}
\usepackage{threeparttable}

\begin{document}
\title{An Additional Resonance Damping Control \\ for Grey-Box D-PMSG Wind Farm Integrated \\ Weak Grid}

\author{Tao Zhang, \emph{ Student Member}, \emph{IEEE}, Songhao Yang, \emph{Member}, \emph{IEEE}, Zhiguo Hao, \emph{Senior Member}, \emph{ IEEE}, \\ Hongyue Ma, \emph{ Student Member}, \emph{IEEE}, Baohui Zhang, \emph{Fellow}, \emph{IEEE}
	
\thanks{This work was supported by National Key Research and Development Program of China (2022YFB2402700), Technology Project of State Grid Corporation of China (52272222001J).

Tao Zhang, Songhao Yang, Zhiguo Hao, Hongyue Ma, Baohui Zhang are with State Key Laboratory of Electrical Insulation and Power Equipment, Xi'an Jiaotong University, Xi'an, China (e-mail: zhghao@mail.xjtu.edu.cn, songhaoyang@xjtu.edu.cn).}
}

\maketitle

\begin{abstract}
Considerable efforts have been made to address the resonance issue of the Direct-drive Permanent Magnet Synchronous Generator (D-PMSG) wind farm integrated power systems. However, the D-PMSG controller structure and parameters are concealed because of commercial secrecy, thus the target system exhibits grey-box characteristics. The existing resonance damping methods are either unavailable for grey-box systems or economically infeasible, which makes resonance damping of grey-box systems extremely challenging. To address this issue, this paper proposes an Additional Resonance Damping  Control (ARDC) specifically for the grey-box D-PMSG system. This strategy is achieved by incorporating an additional control loop outside the D-PMSG controller. Firstly, the external impedance characteristics are obtained by the frequency sweeping technique offline and then the key parameter of the additional control loop is determined by the Bode-diagram-based method under the worst stability scenario. Once the resonance occurs, the external impedance of the black-box D-PMSG is reshaped online to increase the magnitude stability margin of the system, thus providing effective resonance damping. The ARDC's effectiveness is finally verified in the simulation and controller-hardware-in-the-loop experiment under various operating conditions.
\end{abstract}


\section{Introduction}

\IEEEPARstart{I}{n} recent years, there has been a growing concern regarding the frequent resonance accidents attributed to the D-PMSG wind farm (WF) integrated weak grid \cite{Xie_type4,PS_PMSG_weak,SE_harm,4or30Hz_R}. Significant efforts have been devoted to developing suitable control strategies for mitigating resonance. However, the detailed structure and parameters of D-PMSG controllers are not disclosed by manufacturers due to commercial and technical confidentiality \cite{Black-Box_CIG,Gray-Box_Wind}, thereby leaving operators with a grey-box scenario. This complicates the resolution of resonance issues stemming from the integration of the D-PMSG wind farm with the weak grid.

On the issue of resonance damping for the white-box systems, the mitigation methods can be divided into two categories, namely 1) utilizing external equipment \cite{equip_FACTS,wide_suppress_statcom,suppress_statcom,equip_2, Damping_Controller} and 2) improving the wind turbines(WTs)' control strategies \cite{suppress_CSEE,MSC_sup,PCC_com,q_axis_voltage,PLL_yang,IPLL_Li}.

 The external equipment of the former methods include existing equipment  \cite{equip_FACTS,wide_suppress_statcom,suppress_statcom} and additional specialized equipment\cite{equip_2, Damping_Controller}. In ref. \cite{equip_FACTS}, the subsynchronous resonance damping of fixed-speed WTs with series FACTS devices is studied. A designed damping controller is added to the power control loop of the gate-controlled series capacitor (GCSC) to suppress the resonance. Ref. \cite{wide_suppress_statcom} proposes a wideband harmonic voltage feedforward control strategy of static synchronous compensator (STATCOM) to mitigate resonance of the D-PMSG WF. However, the countermeasures based on such FACTS devices are economically infeasible and can only be adopted if such expensive equipment is already installed. Moreover, ref. \cite{equip_2} points out that resonance damping is effective only when the capacity of FACTS equipment is more than 50\% of the installed WTs' capacity. The other utilizing-external-equipment methods adopt a special-purpose shunt-converter concept \cite{equip_2, Damping_Controller}, which is supplemented by an adaptive control to provide multiple-frequency reference currents for injection to achieve active resonance damping. In addition, the methods can online capture the resonant frequency with the aid of a frequency estimator. The special-purpose devices could be a competitive alternative despite the additional investment and the complexity of implementation.

In contrast, the resonance damping methods based on improving WTs' control have received more attention because of their better economy \cite{Miti_Control_better}. For resonance problems in different frequency bands, ref. \cite{suppress_CSEE} proposes an impedance reshaping control strategy based on a combination of active damping and virtual admittance to improve the stability margin of D-PMSG WTs connected to the weak grid. However, the influence of the machine-side converter (MSC) is ignored in the strategy development process, which may affect the damping effects. Ref. \cite{PLL_yang} finds that PLL parameters are critical to the resonance issue, and thus designs a PLL-based resonance damping controller to suppress resonance. Nevertheless, this method requires an accurate estimation of the resonance frequency. More importantly, the above improving-WTs-control-based methods are not available in grey-box D-PMSG wind farms that contain black-box WTs.

 For the resonance issue of the grey-box system, the stability analysis method based on the impedance-/admittance- method can be perfectly applied \cite{Black-Box_Impedance-Based,Argu}, while the resonance damping studies are still inadequate. Among the damping methods described above, directly improving the WTs’ control does not apply to the grey-box systems, since the modification authority of the black-box controller is not available. The external-equipment-based approaches may be suitable for grey-box systems, but they are economically infeasible and subject to planning constraints.

Some attempts have been made to damp the resonance of grey-box systems.  In ref. \cite{Gray-Box_Wind}, the controller parameters of the wind energy conversion system are estimated. Then one can identify which part of the controller has a significant impact on the system stability and reshape the device's impedance accordingly. Such an idea is consistent with the work of ref. \cite{Gray-Box_Reshaping}. However, the above two methods make some assumptions about the control structure of the black-box wind turbine, which requires prior knowledge of the WTs' controller. 

To address the resonance issue of D-PMSG WF integrated weak grid, this paper proposes an Additional Resonance Damping Control (ARDC) for grey-box D-PMSG. The contributions can be summarized as follows:

\hangafter 1
\hangindent 1.5em
\noindent
1) A resonance damping method is proposed for the grey-box D-PMSG system. This is achieved by adding a control loop outside the D-PMSG controller. It enables the impedance reshaping of the grey-box system without the need for expensive external equipment and does not rely on the detailed D-PMSG controller model. The method strikes a favorable balance between economy and applicability.

\hangafter 1
\hangindent 1.5em
\noindent
2) The proposed ARDC can be easily tuned and entirely applicable to the black-box D-PMSG. Only one key parameter needs to be tuned in advance. Once the resonance occurs, the external impedance can be reshaped online by employing the resonance component extraction technique.

\hangafter 1
\hangindent 1.5em
\noindent
3) The proposed method is robust and doesn't negatively impact the target system. The parameter tuning is based on the worst stability condition, thus this method can effectively damp the resonance under various conditions. Besides, the proposed activation criterion prevents ARDC's false start-up in fault conditions. The simulation and controller-hardware-in-the-loop (CHIL) tests eventually verify the efficiency and adaptability of the method. 

The rest of this paper is organized as follows. Section II presents resonance analysis of the grey-box D-PMSG WF integrated weak grid. Section III presents the proposed ARDC. The method is verified through the simulation and the CHIL experiments in Section IV and Section V, respectively. Section VI conducts control verification under faults and the method comparison. Section VII concludes this paper.

\begin{figure}[t]
	\centering
	\subfigure[The D-PMSG grid-tied system's structure.]{
		\includegraphics[width=0.98\linewidth]{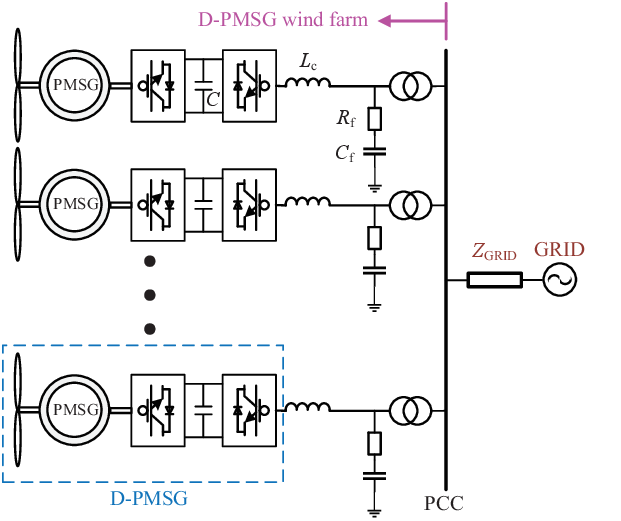} 
		\label{PMSG_grid}
		
	}
	\subfigure[Equivalent block diagram of the D-PMSG grid-connected system.]{
		\includegraphics[width=0.75\linewidth]{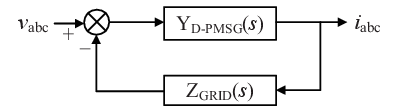} 
		\label{reshape_equ_dia0}
	}
	\DeclareGraphicsExtensions.
	\caption{The target system and its equivalent block diagram.}
	\label{figure1}
\end{figure}
\begin{table}[b]
	\begin{center}
		\caption{The relevant parameters of the D-PMSG grid-tied system}
		\label{PMSG_parameter}
		\begin{tabular}{ c  c  c }
			\toprule
			Parameter & Symbol & Value\\
			\midrule
			D-PMSG Rated Voltage & $ U_{1} $ & 0.69kV\\
			D-PMSG Rated Power & $ S $ & 1.632MVA \\ 
			DC Rated Voltage & $ U_{\rm{dc}} $ & 1.2kV\\
			Rated Frequency & $f_{1}$ & 50Hz \\
			DC capacitor & $C $ & 0.15F\\
			PLL Proportional Gain & $ K_{p p} $ & 0.085 
			\\
			PLL Integral Gain & $ K_{p i} $ & 32 \\
			GSC Proportional Gain
			& $ K_{p} $ & 0.25 
			\\
			GSC Integral Gain
			& $ K_{i} $ & 355 
			\\
			GSC Decoupling Gain & $ K_{r d} $ & 0.0471 
			\\
			MSC Proportional Gain
			& $ K_{mp} $ & 0.8
			\\
			MSC Integral Gain
			& $ K_{mi} $ & 50 
			\\
			MSC Decoupling Gain & $ K_{mr d} $ & 0.2561 
			\\
			DC outer-loop Proportional Gain
			& $ K_{dcp} $ & 0.5 
			\\
			DC outer-loop Integral Gain & $ K_{dci} $ & 5 
			\\
			Connection Inductance & $ L_{\rm{c}} $ &  0.00015H 
			\\
			Filter Resistance & $ R_{\rm{f}} $ &  0.02$ \Omega  $
			\\
			Filter Capacitance & $ C_{\rm{f}} $ &  5.0e-4 F
			
			\\
			Equivalent resistance of the line& $ R_{\rm{Line}} $ & 0.012$ \Omega $ 
			\\
			Equivalent reactance of the line& $ X_{\rm{Line}} $ & 0.25$ \Omega $  
			\\
			\bottomrule
		\end{tabular}
	\end{center}
\end{table}

\section{Resonance Analysis of the Grey-box D-PMSG Integrated Weak Grid}
\label{2}
A typical D-PMSG grid-tied system's structure is shown in Fig. \ref{PMSG_grid}, and the relevant parameters are given in TABLE \ref{PMSG_parameter}. This system will be used as the target system for resonance analysis in Sections \ref{2}, \ref{3}, and \ref{4}.
\subsection{Model Building}
\label{2-1}

 The target system can be split into two subsystems originating from the Point of Common Coupling (PCC), specifically the D-PMSG wind farm (WF) and the AC power grid. By consolidating the entire WF into a single D-PMSG \cite{suppress_statcom, Damping_Controller}, the impedance properties of a single D-PMSG can mirror those of the entire WF. Subsequently, the stability of the D-PMSG WF integrated weak grid can be assessed. The grid can be represented as a series combination of a voltage source and an impedance, while the D-PMSG WF can be represented as a parallel connection of a current source and an admittance \cite{PLL_yang}. Consequently, the target system is simplified as illustrated in Fig. \ref{reshape_equ_dia0}, where $ {\rm{Y}}_{{\rm{D-PMSG}}}(s) $ and $ {\rm{Z}}_{{\rm{GRID}}}(s) $ denote the equivalent admittance of the D-PMSG WF and the impedance of the AC grid, respectively. Leveraging impedance-based methodologies, the stability of the target system can be appraised using the generalized Nyquist criterion or Bode-diagram-based criterion.

Considering the frequency coupling effects of the D-PMSG \cite{fre_coupling}, the MIMO model of the D-PMSG and the AC grid can be denoted as
\begin{equation}
{{\bf{Y}}_{{\rm{D-PMSG}}}}(s) = \left[ {\begin{array}{*{20}{c}}
		{{\rm{Y}}_{{\rm{D-PMSG}}}^{11}(s)}&{{\rm{Y}}_{{\rm{D-PMSG}}}^{12}(s)} \vspace{1ex} \\ 
		{{\rm{Y}}_{{\rm{D-PMSG}}}^{21}(s)}&{{\rm{Y}}_{{\rm{D-PMSG}}}^{22}(s)}
\end{array}} \right]
	\label{PMSG_MIMO_model}
\end{equation}
and
\begin{equation}
{{\bf{Z}}_{{\rm{GRID}}}}(s) = \left[ {\begin{array}{*{20}{c}}
		{{\rm{Z}}_{{\rm{GRID}}}^{1}(s)}&0 \vspace{1ex} \\
		0&{{\rm{Z}}_{{\rm{GRID}}}^{2}(s)}
\end{array}} \right],
	\label{GRID_MIMO_model}
\end{equation}
 respectively. The non-diagonal elements of $ {{\bf{Y}}_{{\rm{D-PMSG}}}}(s) $  indicate the frequency coupling effect. In Eqs. (\ref{PMSG_MIMO_model}) and (\ref{GRID_MIMO_model}), $ s $ is the Laplace operator. When the transfer function is discrete, $ s= j\omega $, and $ \omega $ represents the angular frequency.
 	
 For the D-PMSG's admittance TF matrix  $ {{\bf{Y}}_{{\rm{D-PMSG}}}}(s) $, the specific parameters can be found in works \cite{detail_model,Im_PMSG} and will not be reiterated here. Additionally, it is worth noting that:

\begin{itemize}
\item[1)]

The study focuses on the D-PMSG containing a black-box controller, whose structure and parameters are unknown because of commercial and technical confidentiality \cite{Black-Box_CIG,Gray-Box_Wind}, thus the parameters of  $ {{\bf{Y}}_{{\rm{D-PMSG}}}}(s) $  are unknown and not mandatory;
	
\item[2)]
	
The parameters of $ {{\bf{Y}}_{{\rm{D-PMSG}}}}(s) $  are considered unavailable when performing resonance analysis and control design. a) For the simulation, a pseudo-black-box controller is employed because the parameters are mandatory for simulation modeling. However, the resonance analysis and control do not depend on these specific parameters. b) For the CHIL experiment, a real black-box controller is employed. which is an engineerable controller and one does not have access to its exact structure and parameters.
	
\end{itemize}

For the grid's impedance TF matrix $ {{\bf{Z}}_{{\rm{GRID}}}}(s) $, 
	\[\begin{array}{l}
		{\rm{Z}}_{{\rm{GRID}}}^1(s) = {R_{{\rm{grid}}}} + s{L_{{\rm{grid}}}}\\
		{\rm{Z}}_{{\rm{GRID}}}^2(s) = {R_{{\rm{grid}}}} + {s_2}{L_{{\rm{grid}}}}
	\end{array}\]
where $ {s_2} = s - 2j{\omega _1} $, and $ \omega_1 $ is the fundamental angular frequency; $ {R_{{\rm{grid}}}} $ and $ {L_{{\rm{grid}}}} $ represent the equivalent resistance and inductance of the grid, respectively.

 To analyze the stability more intuitively, the MIMO model $ {\bf{Y}}_{{\rm{D-PMSG}}}(s) $ can be transformed into the equivalent SISO models $ {\rm{Y}}_{{\rm{D-PMSG}}}^1(s) $ and $ {\rm{Y}}_{{\rm{D-PMSG}}}^2(s) $ accounting for the coupling terms according to ref. \cite{MIMO2SISO}, and the detailed transforming method can be written as
 \begin{equation}
 	{\rm{Y}}_{{\rm{D-PMSG}}}^1(s) = {\rm{Y}}_{{\rm{D-PMSG}}}^{11}(s) - \frac{{{\rm{Y}}_{{\rm{D-PMSG}}}^{21}(s){\rm{Y}}_{{\rm{D-PMSG}}}^{12}(s)}}{{{\rm{Y}}_{{\rm{D-PMSG}}}^{22}(s) + {\rm{Y}}_{{\rm{GRID}}}^{2}(s)}},
 	\label{PMSG_SISO_model1}
 \end{equation}
 where 
 $ {\rm{Y}}_{{\rm{GRID}}}^{2}(s) = {[{\rm{Z}}_{{\rm{GRID}}}^{2}(s)]^{ - 1}} $, and
 \begin{equation}
 	{\rm{Y}}_{{\rm{D-PMSG}}}^2(s) = {\rm{Y}}_{{\rm{D-PMSG}}}^{22}(s) - \frac{{{\rm{Y}}_{{\rm{D-PMSG}}}^{12}(s){\rm{Y}}_{{\rm{D-PMSG}}}^{21}(s)}}{{{\rm{Y}}_{{\rm{D-PMSG}}}^{11}(s) + {\rm{Y}}_{{\rm{GRID}}}^{1}(s)}},
 	\label{PMSG_SISO_model2}
 \end{equation}
 where $ {\rm{Y}}_{{\rm{GRID}}}^{1}(s) = {[{\rm{Z}}_{{\rm{GRID}}}^{1}(s)]^{ - 1}} $. After the model transformation, the resonance stability of the system can be directly analyzed by the Bode-diagram-based method.
 \begin{figure}[t]
	\centering
	\subfigure[]{
		\includegraphics[width=0.85\linewidth]{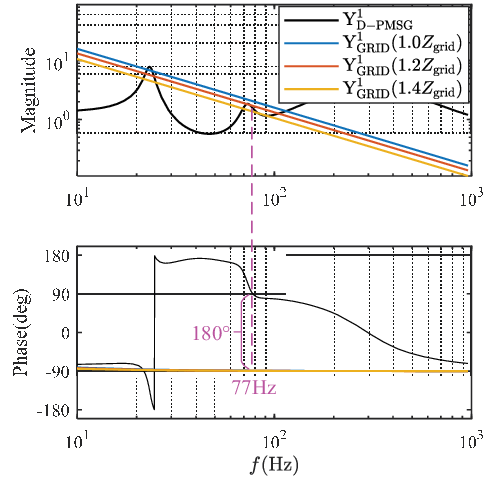} 
		\label{simu_bode1}
		
	}
	\subfigure[]{
		\includegraphics[width=0.85\linewidth]{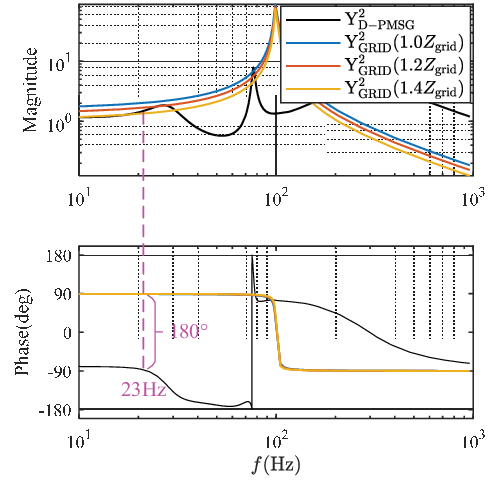} 
		\label{simu_bode2}
	}
	\DeclareGraphicsExtensions.
	\caption{Bode diagrams of the equivalent SISO subsystems.}
	\label{simu_bode}
\end{figure}
It is worth noting that if the target D-PMSG is a black box, its MIMO model can be obtained by frequency sweeping (FS) \cite{Black-Box_Impedance-Based}, and then transformed into two SISO models by Eqs. (\ref{PMSG_SISO_model1}) and (\ref{PMSG_SISO_model2}).  
 
\subsection{Stability Analysis of the Grey-Box System}
\label{2-2}

With the decrease in grid strength, the stability of the D-PMSG grid-tied system weakens. The grid impedance in Fig. \ref{figure1} is set to $ 1.0Z_{\rm{grid}} $, $ 1.2Z_{\rm{grid}} $, and $ 1.4Z_{\rm{grid}} $ respectively, where $ Z_{\rm{grid}}=0.012+j0.25 \Omega  $.

 Fig. \ref{simu_bode} displays the Bode diagrams of the target D-PMSG system, including two sub-graphs, based on the equivalent SISO models. If and only if the two sub-Bode diagrams meet the stability conditions, the target system is stable. 
\begin{figure}[t]
	\centering
	\includegraphics[width=0.98\linewidth]{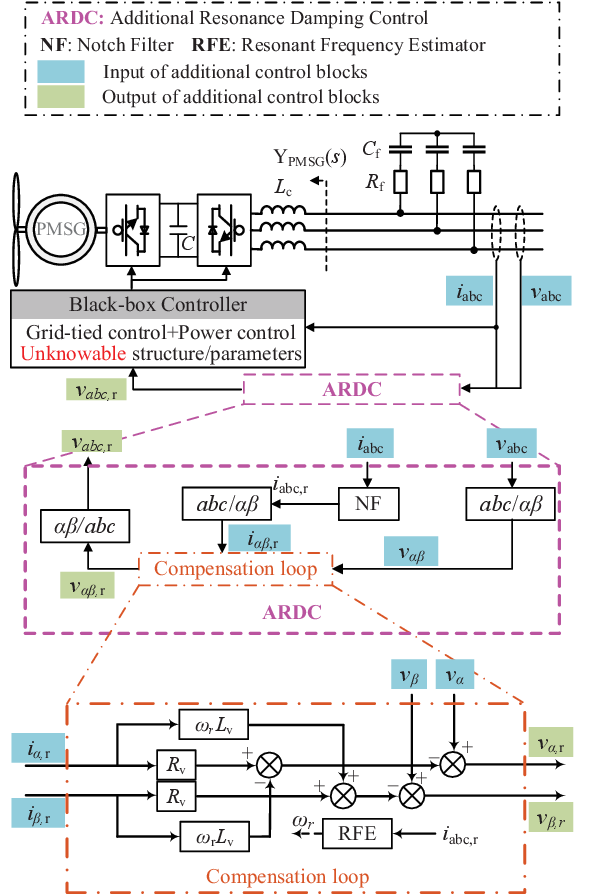}
	\caption{Schematic diagram of the additional control strategy for damping resonance of the grey-box D-PMSG grid-tied system.}
	\label{ARDC}
\end{figure}

One can see that with the increase of the grid impedance, the magnitude of $ {\rm{Y}}_{{\rm{GRID}}}^{1}(s) $ and $ {\rm{Y}}_{{\rm{GRID}}}^{2}(s) $ decrease obviously. When the grid impedance is $ 1.4Z_{\rm{grid}} $, the magnitude of $ {\rm{Y}}_{{\rm{D-PMSG}}}^{1}(s) $ is greater than $ {\rm{Y}}_{{\rm{GRID}}}^{1}(s) $ at 77Hz, where the phase difference between the two is 180°, as shown in Fig. \ref{simu_bode1}. Meanwhile, the magnitude of $ {\rm{Y}}_{{\rm{D-PMSG}}}^{2}(s) $ is greater than $ {\rm{Y}}_{{\rm{GRID}}}^{2}(s) $ at 23Hz, where the phase difference between the two is 180°, as shown in Fig. \ref{simu_bode2}. This suggests that the system is unstable when the grid impedance changes to $ 1.4Z_{\rm{grid}} $, and the potential resonance frequencies are 23/77Hz. When the grid impedance is $ 1.0Z_{\rm{grid}} $ and $ 1.2Z_{\rm{grid}} $, the magnitude of $ {\rm{Y}}_{{\rm{D-PMSG}}}(s) $ is smaller than that of $ {\rm{Y}}_{{\rm{GRID}}}(s) $ at the frequency with a phase difference of 180°, indicating the system is stable.

 Under the weak grid condition, the magnitude margin of the target system decreases, which damages the system's stability. It is necessary to reshape the D-PMSG's impedance to increase the magnitude margin of the system under weak grid conditions, so as to damp the resonance. For this issue, most of the impedance reshaping methods are based on the white-box model \cite{suppress_CSEE,MSC_sup,PCC_com,q_axis_voltage,PLL_yang,IPLL_Li}, and the corresponding controller is allowed to be modified. However, the existence of commercial barriers makes the D-PMSG controller only be regarded as a black box in the actual situation \cite{Black-Box_CIG,Gray-Box_Wind}. To break through this limitation, a novel additional resonance damping control (ARDC) for the grey-box D-PMSG system is designed. It is worth noting that the ARDC is customized for typical PMSG configurations, specifically those connected to the power grid through back-to-back converters, referred to as D-PMSG. The modeling, control design, and validation of control effectiveness in the grey-box system all take into account the complete D-PMSG architecture.

\section{The Proposed ARDC Method}
\label{3}
\subsection{ARDC Design}

A novel ARDC is designed for the grey-box D-PMSG outside the controller. Fig. \ref{ARDC} shows the detailed ARDC structure, which is mainly composed of a notch filter (NF) and a compensation loop.

In Fig. \ref{ARDC}, the inputs to the original controller of the D-PMSG are the voltage and current signals from the grid-tied point, while the outputs are the control signals for the back-to-back converters. For the controller, one can only know that its basic functions are 1) locking the grid angle for grid connection and 2) controlling the WT's power, but its detailed structure and parameters are unknown and inaccessible due to commercial and technical confidentiality \cite{Black-Box_CIG,Gray-Box_Wind,Black-Box_Impedance-Based}, rendering the involved controllers as black-box controllers.

Situated outside the controller, the ARDC is capable of achieving impedance reshaping by processing the voltage signals input to the original converter controller. The notch filter (NF) in the ARDC serves to eliminate the influence of the proposed control on the fundamental component, whereas the resonant frequency estimator (RFE) is utilized for the online parameter tuning of the resonant frequency point. Specifically, the RFE outputs the resonant angular frequency $ \omega_r $ \cite{PLL_yang}. Further details regarding the remaining control loops in the ARDC are provided subsequently.

 The transfer function of NF is 
\begin{equation}
	{{\rm{H}}_{{\rm{NF}}}}(s) = \frac{{{G_0}({s^2} + \omega _0^2)}}{{{s^2} + 2\xi {\omega _0}s + \omega _0^2}},
	\label{}
\end{equation}
where $ G_0 $ and $ \xi $  are the gain and damping coefficients. The effect of the compensation loop on the signal can be expressed as
\begin{equation}
{v_{abc ,{\rm{r}}}} = {v_{abc}} + {i_{abc}}{\rm{H}}_{{\rm{FB}}},
	\label{feed_function}
\end{equation}
where $  {v_{abc}} $  and $ {i_{abc}} $ are the initial voltage and current signals, respectively; $ {v_{abc ,{\rm{r}}}} $ represents the processed voltage signal; and
\begin{equation}
	{\rm{H}}_{{\rm{FB}}} =  - ({R_{\rm{v}}} + j{\omega _r}{L_{\rm{v}}}),
	\label{feed_abc}
\end{equation}
where $ R_{\rm{v}} $ and $ L_{\rm{v}} $  are the introduced virtual resistor and inductor, respectively.

To realize Eq. (\ref{feed_abc}) outside the controller, the voltage $ {v_{{\rm{abc}}}} $  and current $ {i_{{\rm{abc,r}}}} $ are converted from the $abc$ coordinate to the $ \alpha \beta $ coordinate through the Clark transformation. The transfer function of $ 	{\rm{H}}_{{\rm{FB}}} $ in the  $ \alpha \beta $  coordinate can be written as
\begin{equation}
	{\rm{H}}_{{\rm{FB}}}^{\alpha \beta } = \left[ {\begin{array}{*{20}{c}}
			{ - {R_{\rm{v}}}}&{{\omega _r}{L_{\rm{v}}}}\\
			{ - {\omega _r}{L_{\rm{v}}}}&{ - {R_{\rm{v}}}}
	\end{array}} \right].
	\label{}
\end{equation}
\begin{figure}[t]
	\centering
	\includegraphics[width=0.75\linewidth]{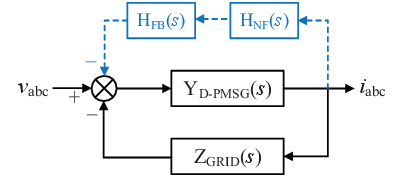}
	\caption{Equivalent block diagram of the D-PMSG grid-connected system with ARDC.}
	\label{reshape_equ_dia}
\end{figure}
\begin{figure}[t]
	\centering
	\subfigure[Form 1]{
		\includegraphics[width=0.75\linewidth]{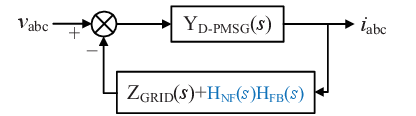} 
		\label{reshape_equ_dia_1}
		
	}
	\subfigure[Form 2]{
		\includegraphics[width=0.75\linewidth]{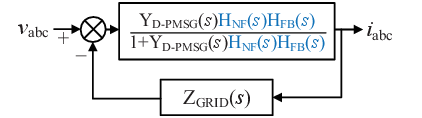} 
		\label{reshape_equ_dia_2}
	}
	\DeclareGraphicsExtensions.
	\caption{Two forms of simplification of the equivalent block diagram.}
	\label{reshape_equ_huajian}
\end{figure}
\begin{figure}[t]
	\centering
	\includegraphics[width=0.85\linewidth]{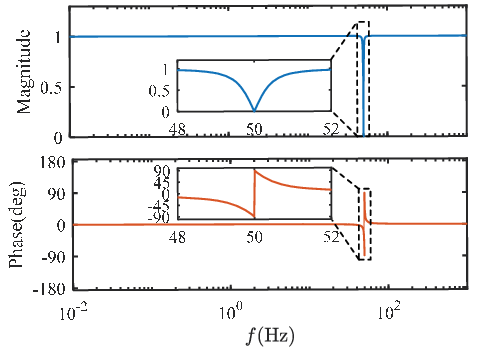}
	\caption{Bode diagram of the designed NF.}
	\label{filter_bode}
\end{figure}
\subsection{ARDC's function}
According to the ARDC's topology, the entire target system's transfer function can be equivalent to the form shown in Fig. \ref{reshape_equ_dia}.  The compensation loop $ {{\rm{H}}_{{\rm{NF}}}}(s)	{\rm{H}}_{{\rm{FB}}} (s) $ is superimposed between the input and output of the original system. According to Mason's formula, the block diagram in Fig. \ref{reshape_equ_dia} can be further simplified to obtain the two forms in Fig. \ref{reshape_equ_huajian}. Form 1 indicates that adopting ARDC is equivalent to adding a supplemental term to the grid impedance. And Form 2 illustrates that utilizing ARDC essentially reshapes the D-PMSG impedance.

From the perspective of Form 1, the grid impedance is reshaped as
\begin{equation}
	{\rm{Z}}_{{\rm{GRID}}}^{\rm{re}}(j{\omega _r}) = {\rm{Z}}_{{\rm{GRID}}}^{}(j{\omega _r}) - ({R_{\rm{v}}} + j{\omega _r}{L_{\rm{v}}}){{\rm{H}}_{{\rm{NF}}}}(s)
	\label{}
\end{equation}
for the resonance components (RCs). This indicates that the proposed ARDC reduces the grid's impedance, thereby improving the stability of the target system. To avoid changing the phase characteristics of the original grid, the power factor angle of the compensation impedance is set to be consistent with that of the grid impedance, i.e., $ {R_{\rm{v}}} = k{R_{{\rm{grid}}}} $ and $ {L_{\rm{v}}} = k{L_{{\rm{grid}}}} $, where $ k $ is the control coefficient.

From the perspective of Form 2, the reshaped D-PMSG's admittance can be represented as
\begin{small}
\begin{equation}
	{\rm{Y}}_{{\rm{D-PMSG}}}^{{\rm{re}}}(s) = \frac{{{\rm{Y}}_{{\rm{D-PMSG}}}^{}(s)}}{{1 - {\rm{Y}}_{{\rm{D-PMSG}}}^{}(s)k{{\rm{H}}_{{\rm{NF}}}}(s)({R_{{\rm{grid}}}} + j{\omega _r}{L_{{\rm{grid}}}})}},
	\label{}
\end{equation}
\end{small}
where $k$ determines the effect of D-PMSG's impedance reshaping. It is noted that $k$ is the only parameter that needs to be tuned in the proposed method.

Fig. \ref{filter_bode} displays the Bode diagram of the designed NF, where $ G_0=1 $ and $ \xi=0.01 $. In the frequency band other than the fundamental frequency (50Hz), it can be considered that the amplitude of NF is 1 and the phase is 0. Therefore, it is believed that $ {{\rm{H}}_{{\rm{NF}}}}(s)=1 $ for the RCs.

\subsection{Parameter Tuning}

For the proposed ARDC, the following parameter tuning principles are adopted. 

1) The equivalent SISO external impedance ($ {\rm{Y}}_{{\rm{D-PMSG}}}^{1}(s) $, $ {\rm{Y}}_{{\rm{D-PMSG}}}^{2}(s) $) of the D-PMSG are obtained by FS offline. It should be noted that the worst stability scenario where WTs have the largest output should be examined for parameter tuning \cite{PLL_yang}. This guarantees the ARDC strategy's efficacy in suppressing resonance under various active power conditions.

2) The control coefficient $ k $ is set online. Assume that the required magnitude margin is $ m  $, then
\begin{equation}
m  = \frac{{{\rm{Y}}_{{\rm{GRID}}}^{}(j{\omega _r})}}{{{\rm{(1 - k)Y}}_{{\rm{D-PMSG}}}^{}(j{\omega _r})}}.
	\label{}
\end{equation}
Therefore, $ k $ can be tuned as
\begin{equation}
k = 1 - \frac{{{\rm{Y}}_{{\rm{GRID}}}^{}(j{\omega _r})}}{{{m\rm{Y}}_{{\rm{D-PMSG}}}^{}(j{\omega _r})}},
	\label{}
\end{equation}
where $ \rm{Y}_{{\rm{D-PMSG}}}^{}(j{\omega _r}) $ is the D-PMSG external characteristic of the resonant point obtained by FS offline and RFE online; $ {{\rm{Y}}_{{\rm{GRID}}}^{}(j{\omega _r})} $ is the grid admittance, which can also be measured online by Quadratic
Reside Binary (QRB) Sequence method \cite{grid_impedance_measure},  then the control coefficient $ k $ can be adjusted online. 

\subsection{ARDC's Activation and Deactivation }
\label{3-D}
\begin{figure}[t]
	\centering
	\subfigure[]{
		\includegraphics[width=0.85\linewidth]{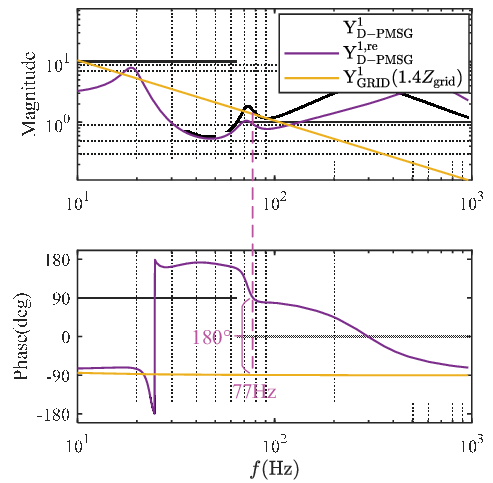} 
		\label{simu_bode1_re}
		
	}
	\subfigure[]{
		\includegraphics[width=0.85\linewidth]{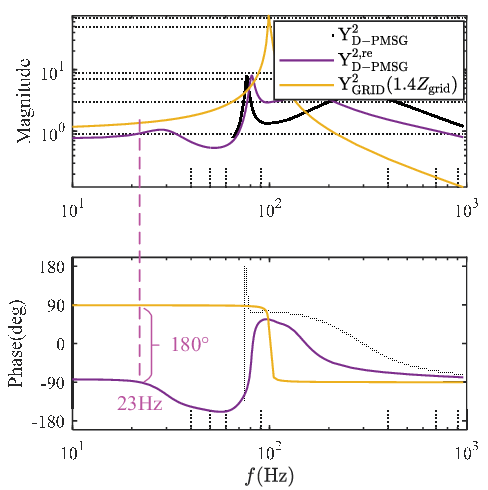} 
		\label{simu_bode2_re}
	}
	\DeclareGraphicsExtensions.
	\caption{Bode diagrams of the target system after reshaping D-PMSG.}
	\label{simu_bode_r}
\end{figure}
\begin{figure}[t]
	\centering
	\includegraphics[width=0.85\linewidth]{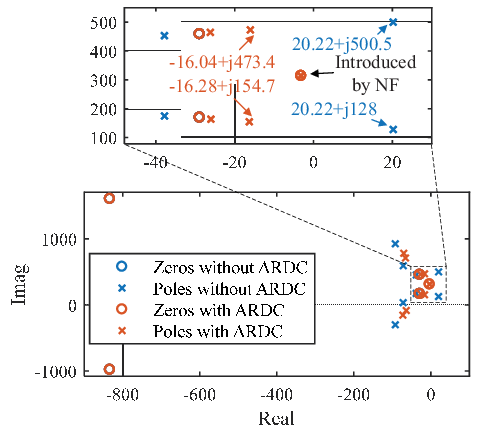}
	\caption{Zero-pole distribution of the target system with/without ARDC.}
	\label{Zero_pole_fina}
\end{figure}

The ARDC is supposed to be activated during resonance and deactivated when the resonance conditions disappear.

 Before activating ARDC, the following situations need to be identified: 1) The system operates stably without RCs; 2) Resonance occurs during the fault but converges  afterward; 3) Resonance occurs and does not converge. 
 ARDC should be activated only in the third situation when RCs' magnitude exceeds the threshold. 
 
The voltages and currents will contain RCs  once resonance occurs. In this paper, the current RC $ i_{\alpha,\rm{r}} $ is selected to reflect the resonance dynamics due to its apparent features. Therefore, the ratio $ r(t) $ of the RCs to the fundamental current component is used as ARDC start-up index, which can be expressed as
\begin{equation}
r(t) = \frac{{{A_r}(t)}}{{{A_0(t)}}},
	\label{}
\end{equation}
where $ A_r(t) $ and $ A_0(t) $ represent the amplitude of the RCs and the fundamental component, respectively. $ r(t) $ exceeding the preset threshold is a necessary condition for ARDC's activation. 

Besides, the ARDC is \emph{blocked} under fault conditions to avoid false activation. Because  the frequency spectrum at the fault stage contains broad-band RCs \cite{PLL_yang}, which may cause $ r(t) $ to exceed the preset threshold. In this paper, the ARDC is blocked when the PCC voltage is below 0.8 p.u..

The frequencies and magnitudes of the RCs in the target system can be obtained by real-time FFT analysis \cite{equip_2}. It is recommended to preset $ \omega_{r} $ as the frequency corresponding to the minimum magnitude margin obtained by the offline FS. Under normal operating conditions, the signal $  i_{\alpha,\rm{r}}  $ is minimal and the ARDC does not work. Once resonance occurs, the index $ r(t) $ exceeds the threshold and the resonance is non-convergent. Then the designed activation criterion is triggered and the ARDC will mitigate the resonance.

On the other hand, the ARDC is deactivated when the resonance cause disappears. The coupling between the grid impedance $ {{\bf{Z}}_{{\rm{GRID}}}}(s) $ and WF admittance $ {{\bf{Y}}_{{\rm{D-PMSG}}}}(s) $ is the direct cause of resonance. The former is white box and can be directly obtained, while the latter can be derived based on offline FS combined with the WF's topology and the WTs' grid-tied information, including the number and location of WTs connected to the grid. Therefore, when the Bode curves calculated by the grid impedance and WF admittance meet the stability requirements, ARDC is deactivated. The method based on SISO models mentioned in Section \ref{2} can be adopted, and will not be repeated here.

\subsection{Performance Analysis}

For the instability situation when the grid impedance changes to $ 1.4Z_{\rm{grid}} $, $k$ is tuned to 0.5 to guarantee that the magnitude margin is larger than 1.4. 

1) Impedance/admittance Reshaping Effect

Bode diagrams of the target system with$/$without ARDC  are shown in Fig. \ref{simu_bode_r}, where $ {\rm{Y}}_{{\rm{D-PMSG}}}^{1}(s) $ and $ {\rm{Y}}_{{\rm{D-PMSG}}}^{2}(s) $ are the D-PMSG admittances without ARDC while $ {\rm{Y}}_{{\rm{D-PMSG}}}^{1,\rm{re}}(s) $ and $ {\rm{Y}}_{{\rm{D-PMSG}}}^{2,\rm{re}}(s) $ are the D-PMSG admittances with ARDC. It can be seen that after adopting ARDC, the phase characteristics of D-PMSG at the resonance points (23Hz and 77Hz) remain unchanged, and the phase difference with the grid is still 180°, while the magnitude characteristics change evidently. Specifically, the magnitude of D-PMSG'admittance at the resonance points is significantly reduced to less than the grid, which changes the magnitude margin of the system to the desired value, meaning that the system stability is effectively enhanced.

2) Closed-Loop Pole Analysis

The small-signal stability of the system can be effectively assessed through the examination of the system eigenvalues. In the context of the analyzed resonance stability concern, it is imperative to note that the system eigenvalues are in alignment with the poles of the system’s closed-loop transfer function \cite{Dominant1,Dominant}. Consequently, the poles of the system’s closed-loop transfer function are computed both with and without the ARDC in order to evaluate the impact of the proposed control strategy on the system's stability.

Without the ARDC, the closed-loop transfer function of the target system can be denoted as
\begin{equation}
	{H_{sys}}(s) = \frac{{{\rm{Y}}_{{\rm{D - PMSG}}}^{}(s)}}{{\rm{I} + {\rm{Y}}_{{\rm{D - PMSG}}}^{}(s)Z_{{\rm{GRID}}}^{}(s)}},
	\label{TF_cls_wuARDC}
\end{equation}
while the system's closed-loop transfer function is
\begin{equation}
	H_{{\rm{sys}}}^{{\rm{re}}}(s) = \frac{{{\rm{Y}}_{{\rm{D - PMSG}}}^{{\rm{re}}}(s)}}{{\rm{I} + {\rm{Y}}_{{\rm{D - PMSG}}}^{{\rm{re}}}(s)Z_{{\rm{GRID}}}^{}(s)}}
	\label{TF_cls_ARDC}
\end{equation}
with the ARDC.

Fig. \ref{Zero_pole_fina} illustrates the zero-pole distribution for $ {H_{sys}}(s) $ and $ H_{{\rm{sys}}}^{{\rm{re}}}(s) $, respectively. It is evident that upon the implementation of ARDC, the dominant poles of the system, which are the poles closest to the imaginary axis, relocate towards the negative half of the real axis. The real part of the original system’s dominant pole measures approximately 22 (with a damping of about -10.9\%). After applying the ARDC to the system, the real part of the system's dominant pole becomes approximately -16 (with a damping of about 10.1\%), signifying a noteworthy enhancement in the system's stability.

\subsection{Characteristics of ARDC}
 
 \begin{figure}[t]
 	\centering
 	\includegraphics[width=0.85\linewidth]{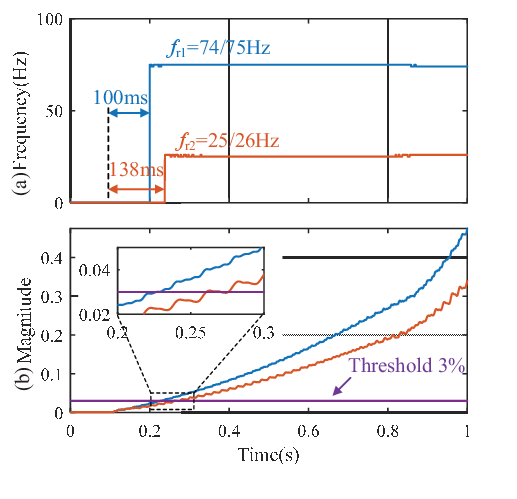}
 	\caption{The frequencies and magnitudes of the RCs in the current.}
 	\label{simulation_fre}
 \end{figure}
 \begin{figure}[t]
 	\centering
 	\subfigure[Active power waveforms]{
 		\includegraphics[width=0.85\linewidth]{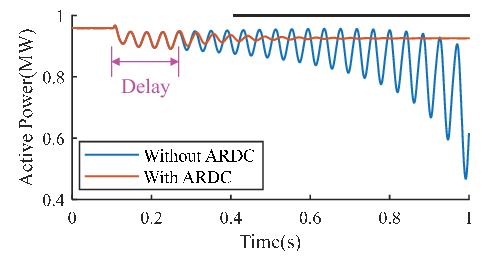} 
 		\label{simu_plot_P}
 	}
 	\subfigure[Resonance current waveforms in the $ \alpha $ axis]{
 		\includegraphics[width=0.85\linewidth]{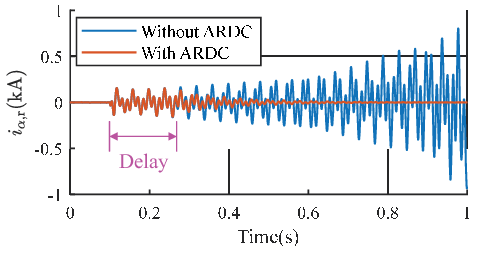} 
 		\label{simu_plot_U}
 	}
 	\DeclareGraphicsExtensions.
 	\caption{The simulation waveforms with$/$without ARDC.}
 	\label{simu_P_and_U}
 \end{figure}

In summary, the proposed ARDC has the following characteristics:

\emph{1) Simple Structure} In ARDC, the additional control loop is structurally simple and contains only a feedback loop. Also, only the control coefficient $ k $ needs to be tuned to achieve the impedance reshaping.

\emph{2) Flexible Adaptability}  The proposed ARDC has flexible adaptability for the black-box D-PMSG. It can be applied to the D-PMSGs with different controller structures and parameters by adding the external control loop outside the controller. 

\emph{3) Strong Robustness} ARDC is robust because it is parameterized based on the worst-condition stability and the activation criterion is set to prevent its false activation.

\section{Simulation Validation}
\label{4}
The effect of the proposed ARDC is first verified by the time-domain simulations. The target D-PMSG grid-tied system is built on the PSCAD/EMTDC simulation platform.

\begin{figure}[t]
	\centering
	\subfigure[Hardware connection]{
		\includegraphics[width=0.95\linewidth]{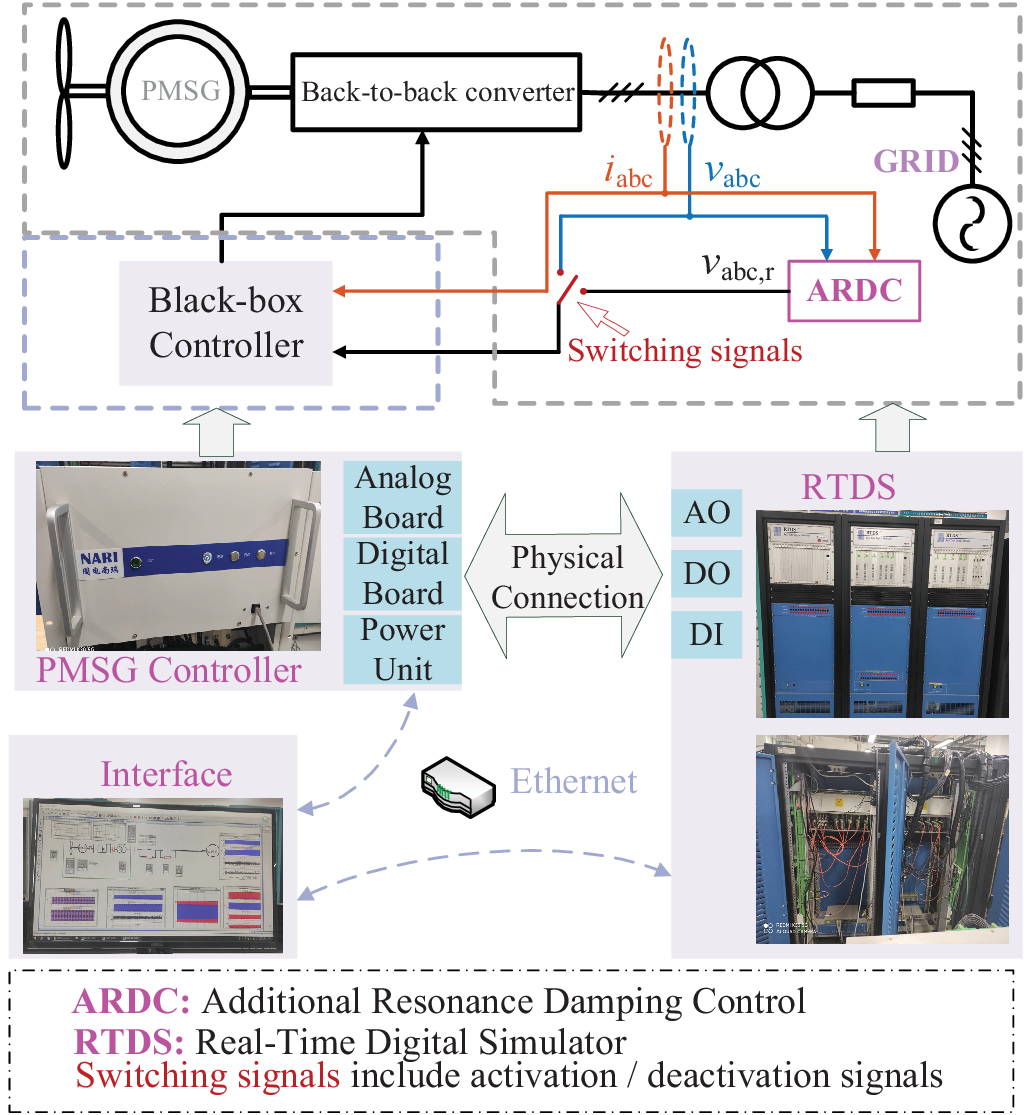} 
		\label{RTDS_1}
	}
	\subfigure[Signal connection]{
		\includegraphics[width=0.95\linewidth]{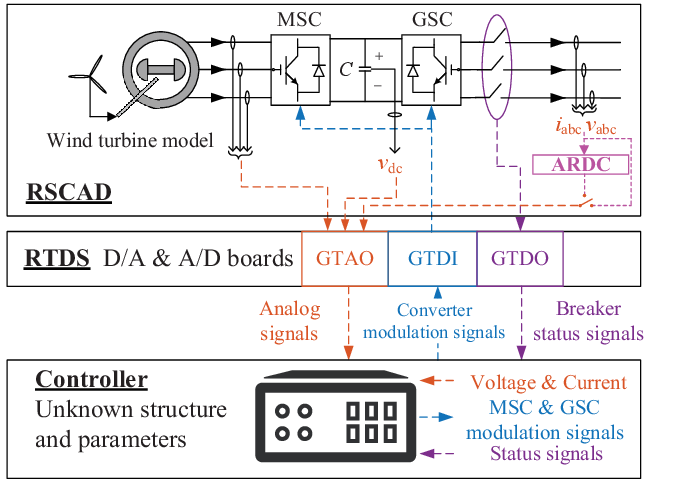} 
		\label{RTDS_2}
	}
	\DeclareGraphicsExtensions.
	\caption{Schematic diagram of the D-PMSG system on the CHIL experimental platform.}
	\label{Damping_RTDs}
\end{figure}
 \begin{table}[b]
	\begin{center}
		\caption{Some parameters of the D-PMSG grid-connected system on RTDS}
		\label{tab_RTDs}
		\begin{tabular}{ c  c  c }
			\toprule
			Parameter & Symbol & Value\\
			\midrule
			Rated voltage of power grid & $ U_{\text{grid}} $ & 35kV\\
			Rated Voltage of D-PMSG  & $ U_{\text{D-PMSG}} $ & 0.69kV\\
			Rated Power & $ S $ & 2.1MVA \\ 
			Rated Frequency & $f_{1}$ & 50Hz \\
			Equivalent Reactance of Transformer & $ Z_\text{T} $ & 0.06p.u. 
			\\
			Grid Resistance& $ {R_{{\rm{grid}}}} $ & 0.011$ \Omega $  \\
			Grid Inductance& $ {L_{{\rm{grid}}}} $ & 0.025H\\
			\bottomrule
		\end{tabular}
	\end{center}
\end{table}

\begin{figure}[t]
	\centering
	\subfigure[]{
		\includegraphics[width=0.85\linewidth]{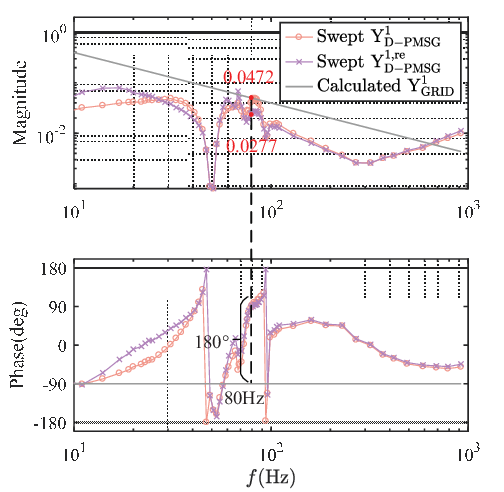} 
		\label{RTDS_Bode_1}
		
	}
	\subfigure[]{
		\includegraphics[width=0.85\linewidth]{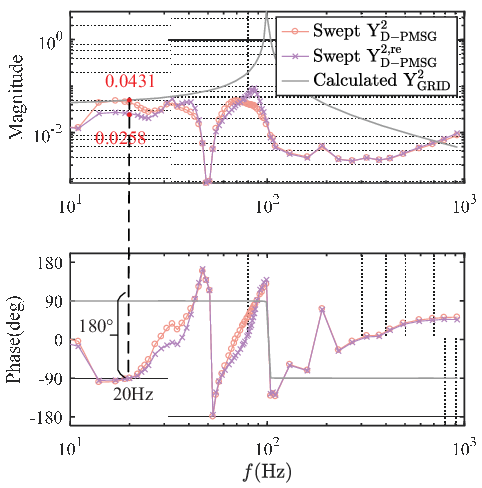} 
		\label{RTDS_Bode_2}
	}
	\DeclareGraphicsExtensions.
	\caption{Bode diagram of the target system without/with ARDC in the CHIL experiment. Circle: without ARDC; Cross: with ARDC.
	}
	\label{RTDS_Bode}
\end{figure}

When the grid impedance is changed from $ 1.2Z_{\rm{grid}} $ to $ 1.4Z_{\rm{grid}} $ ($ Z_{\rm{grid}}=0.0012+j0.025 \Omega  $) at 0.1s, the resonance occurs. The frequencies and magnitudes of the RCs in the current at PCC are shown in Fig. \ref{simulation_fre}. As seen from the figure, the resonance frequencies are 25/26Hz and 74/75Hz, and the magnitude of 74/75Hz resonance is stronger, which is consistent with the theoretical analysis in Subsection \ref{2-2}.

The resonance magnitude trend of the system is shown in Fig. \ref{simulation_fre}(b), and the threshold is set to 3\% of the fundamental component. It can be seen that the resonance magnitude increases after the grid impedance changes, and it exceeds the threshold within 150ms. To ensure that the RFE can output the resonance frequency $ \omega_{r} $, ARDC is set to start up after a delay of 150ms, which is sufficient for online frequency estimation \cite{RFE1,RFE2}.

Fig. \ref{simu_P_and_U} presents the simulation waveforms with$/$without ARDC. The resonance occurs in the system after increasing the grid impedance to $ 1.4Z_{\rm{grid}} $, and the resonance tends to diverge without ARDC. When adopting ARDC, the resonance converges after the delay, and the resonance completely disappears at about 0.7s. It indicates the effectiveness of the proposed ARDC for damping resonance.

\section{Controller Hardware-in-Loop Experiments}
\label{5}

To confirm the effectiveness of the proposed ARDC for the grey-box D-PMSG system, the target system is built on a CHIL experimental platform, with its schematic diagram shown in Fig. \ref{Damping_RTDs}, including the hardware and signal connections. Some parameters are shown in TABLE \ref{tab_RTDs}.

\begin{figure*}[t]
	\centering
	\includegraphics[width=1\linewidth]{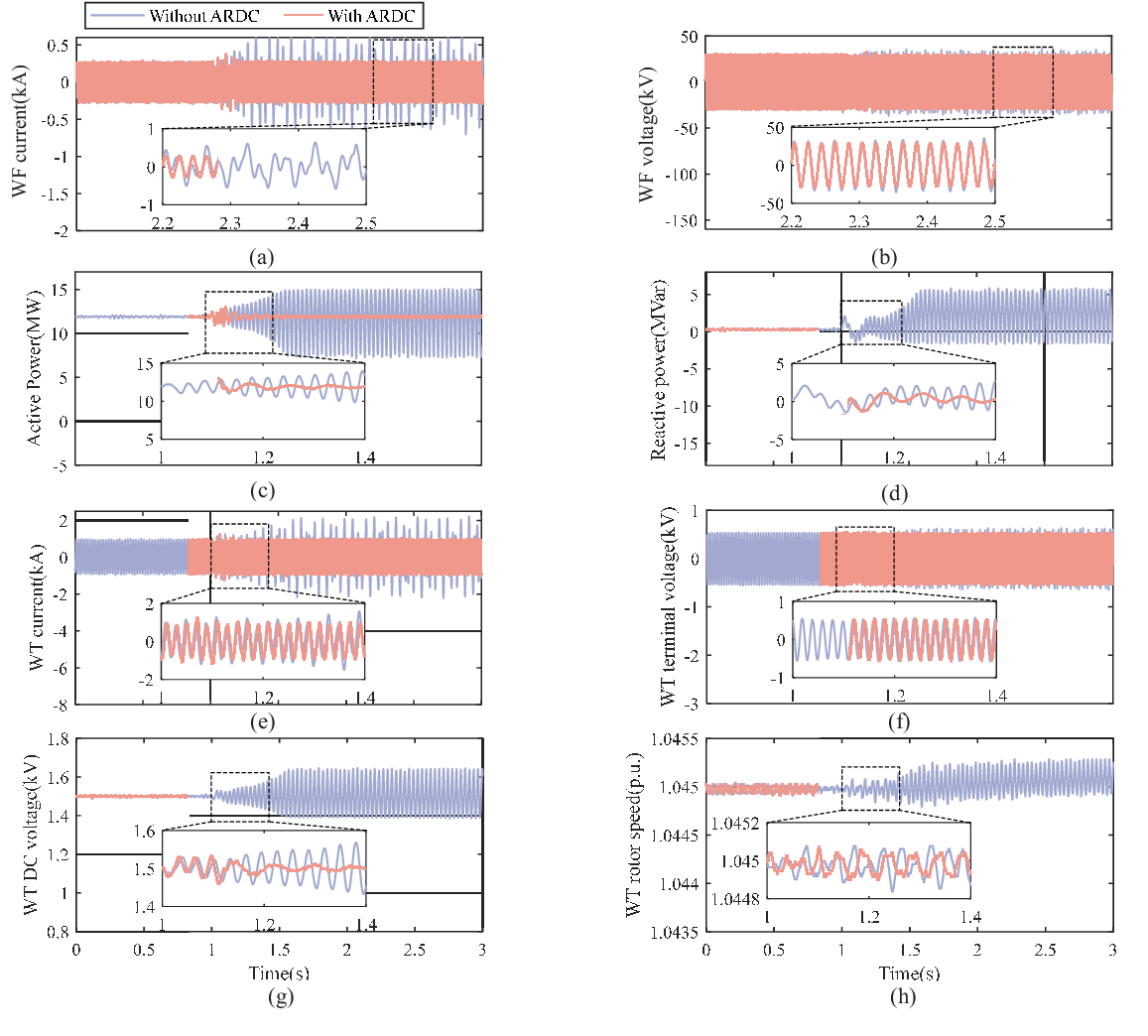}
	\caption{Experimental waveforms of RTDS for increasing grid impedance from $ {Z_{{\rm{grid}}}} $ to $ 1.375{Z_{{\rm{grid}}}} $.}
	\label{RTDS_anli_IUP_zong1}
\end{figure*}

The controller of D-PMSG adopts the real controller produced by NARI Tech. in China, and the controller is a black box. The rest of the target system, such as the wind turbine, the permanent-magnet machine, the transmission line, the transformer, and the ARDC loop, are simulated on the real-time digital simulator (RTDS). 

The RTDS utilized in this study makes use of PB5 processor cards to replicate the power system network on a Real-Time Operating System (RTOS). Simulations are carried out simultaneously across multiple 1.7 GHz PB5 processor cards using RTOS, thereby creating the impression that the power system simulation output is executing in real time \cite{RTDS}. During the CHIL experiment, simulations operate with a standard time step of 50 microseconds for real-time processing. Physical interfacing between the RTDS and the actual D-PMSG controller is facilitated through GTAO, GTDO, and GTDI boards, with software RSCAD responsible for signal configuration and retrieval from the RTDS, as shown in Fig. \ref{RTDS_2}. Additionally, information exchange between the computer, RTDS, and the D-PMSG controller occurs via Ethernet, thereby ensuring the highest possible reliability of the validation results.

The D-PMSG is a black-box device and its external characteristics, including $ {\rm{Y}}_{{\rm{D-PMSG}}}^{11}(s) $, $ {\rm{Y}}_{{\rm{D-PMSG}}}^{12}(s) $, $ {\rm{Y}}_{{\rm{D-PMSG}}}^{21}(s) $ and $ {\rm{Y}}_{{\rm{D-PMSG}}}^{22}(s) $, can be obtained by FS \cite{Black-Box_Impedance-Based}, then the SISO models of the black-box device can be calculated through Eqs. (\ref{PMSG_SISO_model1}) and (\ref{PMSG_SISO_model2}). Since the grid is a white box, one can directly get its external characteristics, $ {\rm{Y}}_{{\rm{GRID}}}^1(s) $ and $ {\rm{Y}}_{{\rm{GRID}}}^2(s) $. Fig. \ref{RTDS_Bode} displays the Bode diagrams of the equivalent SISO models of the system with/without ARDC.

It can be seen that the system without ARDC is in a critical stable state under this condition, and the potential resonant frequency is 20/80Hz. In Fig. \ref{RTDS_Bode_1}, the phase difference at the magnitude intersection of $ {\rm{Y}}_{{\rm{D-PMSG}}}^1(s) $ and  $ {\rm{Y}}_{{\rm{GRID}}}^1(s) $ is exactly 180°, and the corresponding frequency is about 80Hz; in Fig. \ref{RTDS_Bode_2}, when the frequency is 20Hz, the phase difference between $ {\rm{Y}}_{{\rm{D-PMSG}}}^2(s) $ and $ {\rm{Y}}_{{\rm{GRID}}}^2(s) $ is 180°, and the two exactly intersect.

\begin{figure}[t]
	\centering
	\subfigure[Time-domain waveform ]{
		\includegraphics[width=0.85\linewidth]{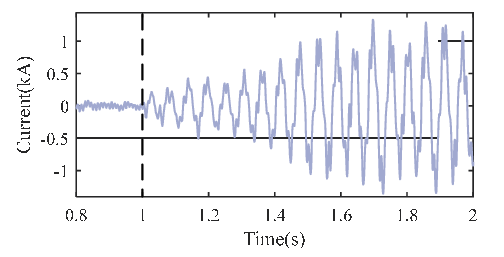} 
		\label{RTDS_anli2_2_I_alpha}
		
	}
	\subfigure[Frequency and Magnitude]{
		\includegraphics[width=0.95\linewidth]{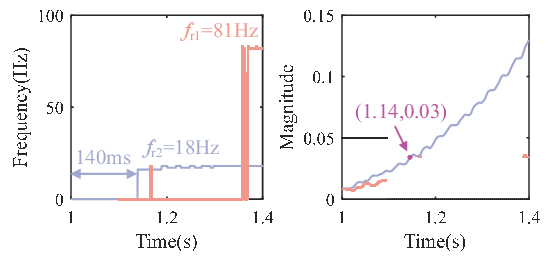} 
		\label{RTDS_anli2_2_start}
	}
	\DeclareGraphicsExtensions.
	\caption{The waveform $ i_{\alpha,\rm{r}}$ in the experiment .}
	\label{RTDS_Ial_sta}
\end{figure}

Set the required magnitude margin as 1.6 and the control coefficient $ k $ can be adjusted based on this. From the swept reshaped model ($ {\rm{Y}}_{{\rm{D-PMSG}}}^{{\rm{1,re}}} $, $ {\rm{Y}}_{{\rm{D-PMSG}}}^{{\rm{2,re}}} $) in Fig. \ref{RTDS_Bode}, the phase characteristics of D-PMSG are almost unaffected after using ARDC, but noticeable changes are observed in the magnitude characteristics. Specifically, the admittance magnitude of the resonant frequency 20/80Hz significantly decreases, resulting in a reshaped magnitude margin of approximately 1.7, which is roughly consistent with the required magnitude margin.

\emph{Scenario 1}:

\begin{figure}[t]
	\centering
	\includegraphics[width=0.95\linewidth]{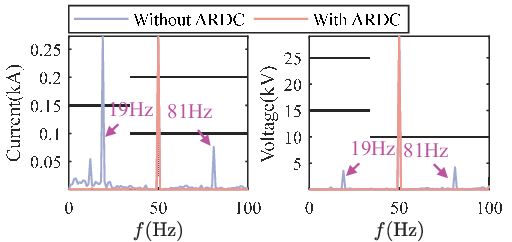}
	\caption{FFT analysis of the experimental waveforms.}
	\label{RTDS_UI_FFT}
\end{figure}

\begin{figure}[t]
	\centering
	\includegraphics[width=0.95\linewidth]{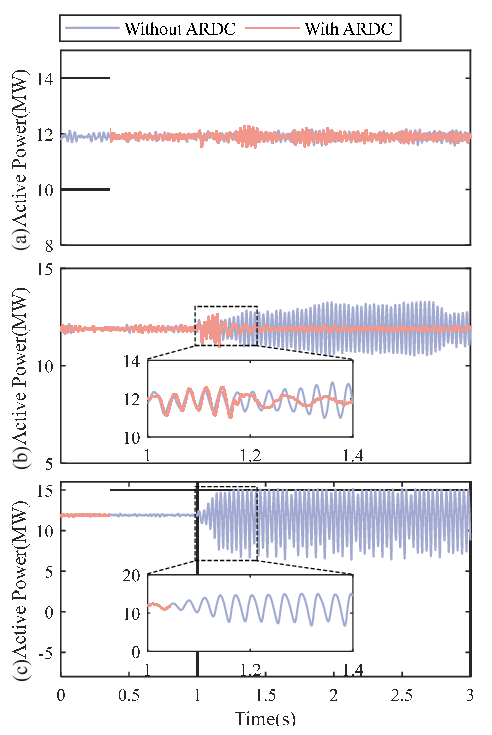}
	\caption{The transmitted active power waveforms of the system with/without ARDC under different grid impedances. (a) $ {Z_{{\rm{grid}}}}$ $ \rightarrow$ 1.1${Z_{{\rm{grid}}}} $; (b) $ {Z_{{\rm{grid}}}}$$ \rightarrow$ 1.25${Z_{{\rm{grid}}}} $; (c) $ {Z_{{\rm{grid}}}}$$ \rightarrow$ 1.5${Z_{{\rm{grid}}}} $.}
	\label{RTDS_anli2}
\end{figure}
Change the grid impedance from $ {Z_{{\rm{grid}}}} $ to $ 1.375{Z_{{\rm{grid}}}} $ at 1s, which means that the short circuit ratio (SCR) of the system has changed from 1.8 to 1.3. 

The waveforms of current, voltage, active power, and reactive power for the WF transmission line are illustrated in Fig. \ref{RTDS_anli_IUP_zong1}(a-d). Meanwhile, the waveforms of the variables relating to the WT itself are depicted in Fig. \ref{RTDS_anli_IUP_zong1}(e-h), including the WT terminal voltage, WT current, WT DC voltage, and WT rotor speed. When the grid impedance increases, significant resonance occurs in the target system without ARDC, and the current resonance is more obvious than the voltage resonance. Correspondingly, the active/reactive power appears to resonate.

 Fig. \ref{RTDS_Ial_sta} displays $ i_{\alpha,\rm{r}}$  during the resonance process, including the waveform in Fig. \ref{RTDS_anli2_2_I_alpha}, and the corresponding frequencies and magnitudes obtained by RFE in Fig. \ref{RTDS_anli2_2_start}. It can be seen that about 140ms after the impedance increases, the RFE identifies that the system resonance frequency is 18 Hz, and the RC's magnitude is about 3\% of the fundamental component, which satisfies the ARDC activation condition. When reaching the constant-magnitude resonance stage, the resonant frequencies of the system is 19/81Hz according to the FFT analysis of the 2.2-2.5s data in Fig. \ref{RTDS_UI_FFT}, which is consistent with the stability analysis results in Fig. \ref{RTDS_Bode}. 

 From Figs. \ref{RTDS_anli_IUP_zong1}-\ref{RTDS_UI_FFT}, it can be seen that the resonance is effectively damped for the target system with the ARDC. Meantime, the FFT analysis results show that the voltage and current do not contain RCs. Note that the resonance converges rapidly after a Delay in Fig. \ref{RTDS_anli_IUP_zong1}(c). The Delay, set to 150ms here, should be greater than the estimation time of the RFE for the RCs. 
 
\emph{Scenario 2}:

At 1s, the grid impedance is adjusted to 1.1$ {Z_{{\rm{grid}}}} $ (SCR=1.63), 1.25$ {Z_{{\rm{grid}}}} $ (SCR=1.44) and 1.5$ {Z_{{\rm{grid}}}} $ (SCR=1.2), respectively. The transmitted active power waveforms of the system with$/$without ARDC are present in Fig. \ref{RTDS_anli2}.

With the increase of grid impedance, different degrees of resonances occur in the target system without ARDC. The grid impedance of 1.1$ {Z_{{\rm{grid}}}} $ does not cause system resonance and ARDC doesn't work.  When the grid impedance further increases, significant resonance occurs. One can see that the greater the grid impedance increases, the larger the amplitude of the system resonance, which indicates that a weaker grid implies poorer stability. In Figs. \ref{RTDS_anli2}(b) and \ref{RTDS_anli2}(c), ARDC can effectively damp the resonance, which demonstrates the effectiveness of the method in damping the resonance under the increasing grid impedance condition.

\begin{figure}[t]
	\centering
	\includegraphics[width=0.95\linewidth]{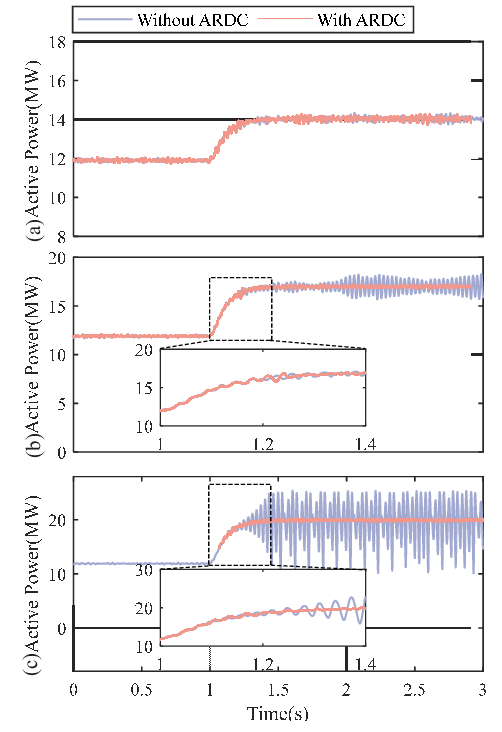}
	\caption{The transmitted active power waveforms of the system with/without ARDC under different D-PMSG grid-tied numbers.(a) 8$ \rightarrow$9; (b) 8$ \rightarrow$11; (c) 8$ \rightarrow$13.}
	\label{RTDS_anli3}
\end{figure}

\emph{Scenario 3}:

The number of grid-connected D-PMSGs is changed from 8 to 9, 11, and 13 respectively at 1s, and the damping effect of the proposed ARDC on resonance under different numbers of grid-connected D-PMSG units is studied. The transmitted active power waveforms with$/$without ARDC are present in Fig. \ref{RTDS_anli3}.

 The system remains stable When the grid-connected D-PMSG number is changed to 9 (SCR=1.6), while resonance occurs when the number is altered to 11 (SCR=1.3) and 13 (SCR=1.1). The latter case has a more pronounced resonance phenomenon. In contrast, in the system containing ARDC, the resonance at the early stage is effectively suppressed, thus avoiding the occurrence of system instability.

In summary, the proposed ARDC can effectively damp the resonance of the grey-box D-PMSG WF integrated into the weak grid in different scenarios.

\section{Discussion}

\label{6}

\subsection{Verification Under Faults}

\begin{figure}[t]
	\centering
	\includegraphics[width=0.9\linewidth]{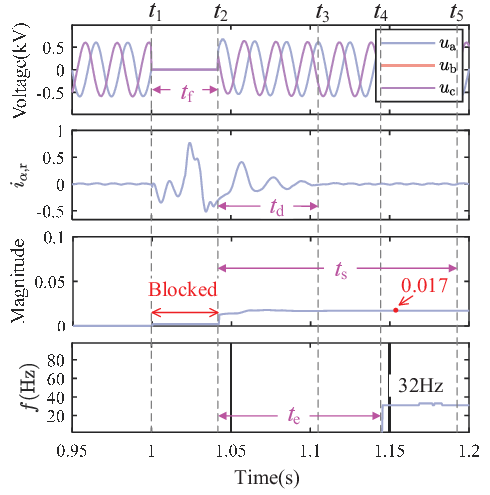}
	\caption{CHIL experimental waveforms during the three-phase fault. $ t_1 $: 1s; $ t_2 $: 1.04s; $ t_3 $: 1.1s; $ t_4 $: 1.146s; $ t_5 $: 1.19s.}
	\label{RTDS_fau}
\end{figure}

The three-phase short-circuit fault condition at the PCC is adopted to verify the robustness of the ARDC's activation.

Based on the analysis in ref. \cite{PLL_yang}, the relevant times and their ranges are listed in TABLE \ref{tab_time}. After a fault occurs, there may be two situations: 1) Relay protection operates within $ t_\text{f} $, which is less than 100 ms. Since ARDC activation time is 150 ms, ARDC will not start up under such scenarios; 2) The system recovers after a transient fault, in which situation the conventional damping scheme is prone to false start. The experimental waveforms in the second situation are shown in Fig. \ref{RTDS_fau}.

\begin{table}[t]
	\begin{center}
		\caption{The relevant times and their ranges}
		\label{tab_time}
		\begin{tabular}{ c  c  c }
			\toprule
			Symbol & Meaning & Range\\
			\midrule
			$ t_\text{f} $  & the time of fault duration & \textless 100ms\\
			$ t_\text{d} $  & the decay time of RCs & related to $\tau $  \tnote{*}\\
			$ t_\text{s} $ &  the start-up time of ARDC & 150ms \\ 
			$ t_\text{e} $ &  the time delay of RFE & \textless150ms \\
			\bottomrule
		\end{tabular}
		\begin{tablenotes}
			\footnotesize
			\item[*] $\tau $ is the system time constant \cite{Argu}.
		\end{tablenotes}
	\end{center}
\end{table}

\emph{Scenario 4}:

\begin{figure}[t]
	\centering
	\includegraphics[width=0.85\linewidth]{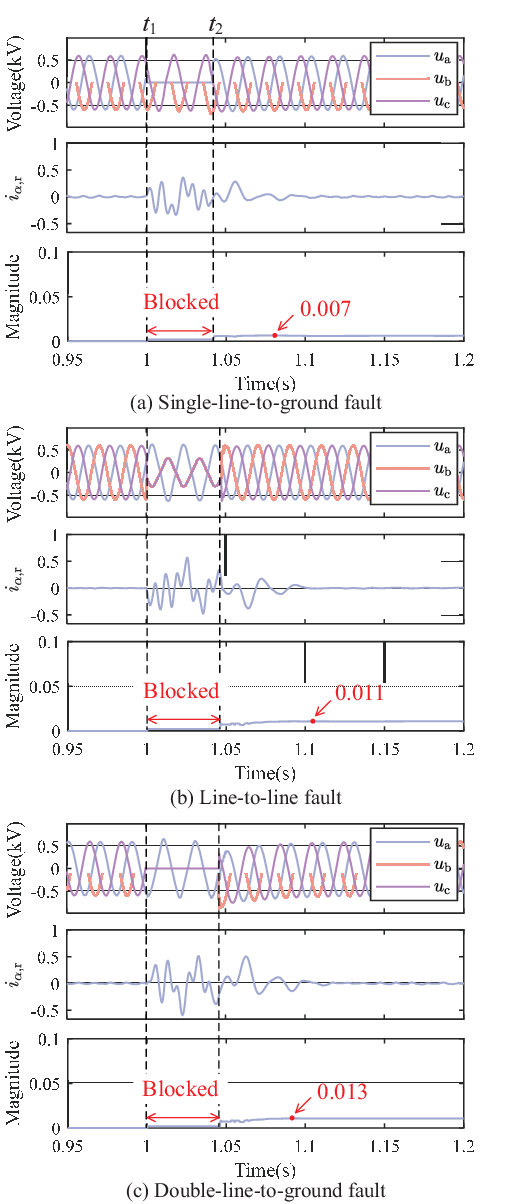}
	\caption{Experimental waveforms under different unbalanced faults. $ t_1 $: 1s; $ t_2 $: 1.04s.}
	\label{Different fault waves}
\end{figure}

A three-phase fault occurs at 1s ($ t_\text{1} $) and the fault duration time is 40ms ($   t_\text{f} $), after which the fault is eliminated. From the $ i_{\alpha,\rm{r}} $ waveform, the decay time of the RCs is 60ms ($ t_\text{d} $). Although the RFE identifies the system as containing the RC of approximately 32Hz after 106ms ($ t_\text{e} $), ARDC will not be activated because the maximum resonance magnitude is 0.017 (less than the threshold value of 0.03). It is worth noting that during the fault period (from $ t_\text{1} $ to $ t_\text{2} $), the ARDC is blocked and the data used for real-time FFT analysis are set to 0.
 
Furthermore, in light of that three-phase faults represent the most severe form of disturbance, the RCs resulting from other faults and perturbations are expected to be below the threshold value. Upon the occurrence of an unbalanced fault, the magnitudes of RCs are observed to be smaller compared to those arising from three-phase faults, as evidenced in Fig. \ref{Different fault waves}. The figure depicts the WT terminal voltage, $ i_{\rm{\alpha,r}} $, and the magnitude of the maximum RC under various unbalanced fault conditions. It indicates that the activation criterion avoids the false activation of ARDC under the fault and perturbation conditions.

 \begin{table*}[t]
	\begin{center}
		\caption{Comparison Between ARDC with the Existing Resonance Damping Methods}
		\label{T4}
		\begin{tabular}{ c  c  c  c  c }
			\toprule 
			Methods & Description& Support grey-box system & No primary equipment required & Support online tuning\\[0.5ex]
			\midrule
		Ref. \cite{MSC_sup} & Grid side compensation control & $ {\times} $   & $ {\surd} $& $ {\times} $ \\ [0.5ex]
			Ref. \cite{IPLL_Li} & A novel phase-locked loop &  $ {\times} $ & $ {\surd} $ &  $ {\surd} $\\[0.5ex]
			Ref. \cite{PLL_yang} & Additional control inside the controller & $ {\times} $ & $ {\surd} $ &  $ {\surd} $\\[0.5ex]
			Refs. \cite{equip_2, Damping_Controller} & Grid-side damping controller & $ {\surd} $  & $ {\times} $ & ${\surd} $  \\[1ex]
			This paper & Additional control outside the controller & $ {\surd} $  & $ {\surd} $ & $ {\surd} $ \\
		\bottomrule
		\end{tabular}
	\end{center}
\end{table*}

\subsection{Method Comparison}

\begin{figure}[t]
	\centering
	\includegraphics[width=0.95\linewidth]{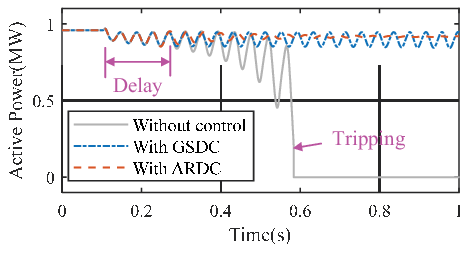}
	\caption{Damping effect comparison of different control methods. GSDC: the method in Refs. \cite{equip_2, Damping_Controller}.}
	\label{sim_com_plot_P}
\end{figure}
The proposed ARDC is compared with the existing resonance damping approaches for D-PMSG systems across three key dimensions: its suitability for grey-box systems incorporating black-box controllers (scenario applicability), the requirement for primary equipment (economy), and its support for online tuning (flexibility). The comparative findings are outlined in TABLE \ref{T4}.

Most of the existing resonance damping methods are found to inadequately address the resonance damping requirements of grey-box D-PMSG grid-tied systems. Notably, the method proposed in refs. \cite{equip_2, Damping_Controller} can apply to the control of grey-box systems by installing a damping controller on the WT grid side for resonance suppression. By comparing the grid side damping controller (GSDC) based method with the proposed ARDC, one can find that GSDC is the control of the WF level, while ARDC is the control of the WT level. Both methods can address the problem that the WTs’ control structure and parameters cannot be obtained in practical engineering, and can enable online parameter tuning to match changes in resonance frequency. Admittedly, the ARDC requires control modifications to the WTs in the target WF, but it is a common limitation of the WT-level resonance control \cite{MSC_sup,PLL_yang,IPLL_Li}.

The distinction between GSDC and ARDC lies in that implementing GSDC necessitates primary equipment, potentially leading to reduced cost-effectiveness. Additionally, due to the limited capacity of GSDC equipment, its ability to dampen resonance is also constrained. Fig. \ref{sim_com_plot_P} illustrates a severe resonance case in a D-PMSG system, where the capacity of GSDC is set at 10\% of the entire WF, which is actually excessive \cite{equip_2}. In this scenario, it is evident that GSDC can only suppress resonance divergence to a certain extent, while the proposed ARDC can successfully suppress the resonance.

\section{Conclusion}

The paper introduces a novel additional resonance damping control (ARDC) designed specifically for the D-PMSG WF integrated weak grid. This control approach achieves effective resonance damping through impedance reshaping of the grey-box D-PMSG containing the black-box controller, which consists of a simple filter and a compensation loop in the $ \alpha\beta $-frame outside the controller.

The ARDC exhibits great effectiveness in suppressing resonance. The parameter tuning is conducted under the worst stability conditions to ensure its effectiveness at various operating points. Experimental results indicate that ARDC significantly increases the stability magnitude margin, from the critical state to 1.7. Even under severe resonance conditions with a low short-circuit ratio of 1.1, the ARDC successfully suppresses system resonance.

The ARDC exhibits favorable robustness against faults. An activation strategy has been designed for the ARDC to prevent incorrect start-up under faults. Experimental results show that, under balanced/unbalanced faults, the ARDC reliably remains inactive and the non-fundamental electrical components during the fault recovery do not trigger the ARDC.

Future explorations will focus on extending the resonance damping strategy to encompass other full-power inverter-based grid-tied systems based on the proposed control.

\balance

\nocite{*}

\bibliographystyle{IEEEtran}
\bibliography{IEEE_EC}

\begin{IEEEbiography}[{\includegraphics[width=1in,height=1.25in,clip,keepaspectratio]{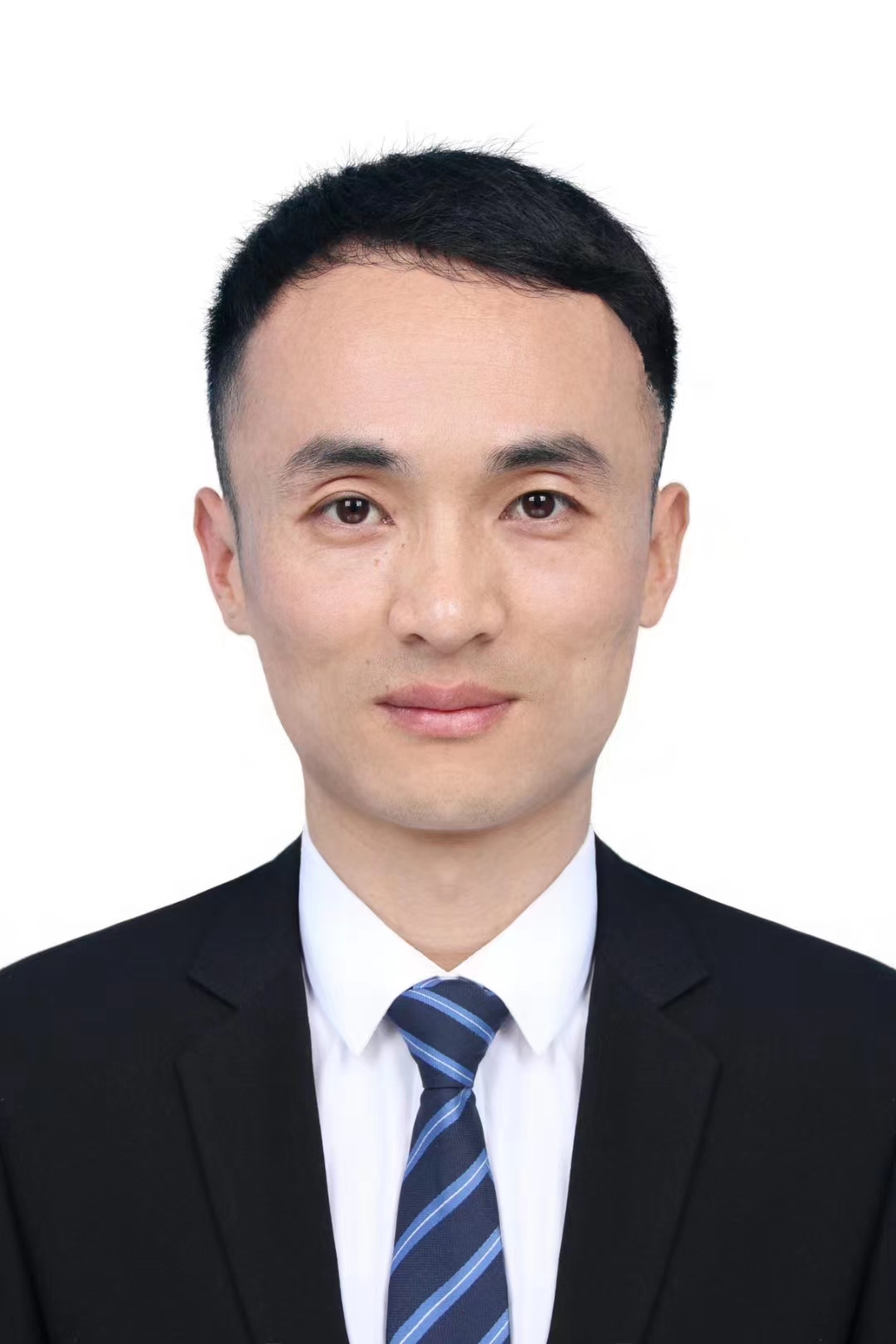}}]{Tao Zhang}
	(S’20) received the B.S. degree in electrical engineering from Xi’an Jiaotong University, Xi’an, China in 2017. And he is currently pursuing a Ph.D. degree on electrical engineering in Xi’an Jiaotong University.  
	
	His research interest includes small-signal stability analysis and control of renewable energy power systems.
\end{IEEEbiography}
\begin{IEEEbiography}[{\includegraphics[width=1in,height=1.25in,clip,keepaspectratio]{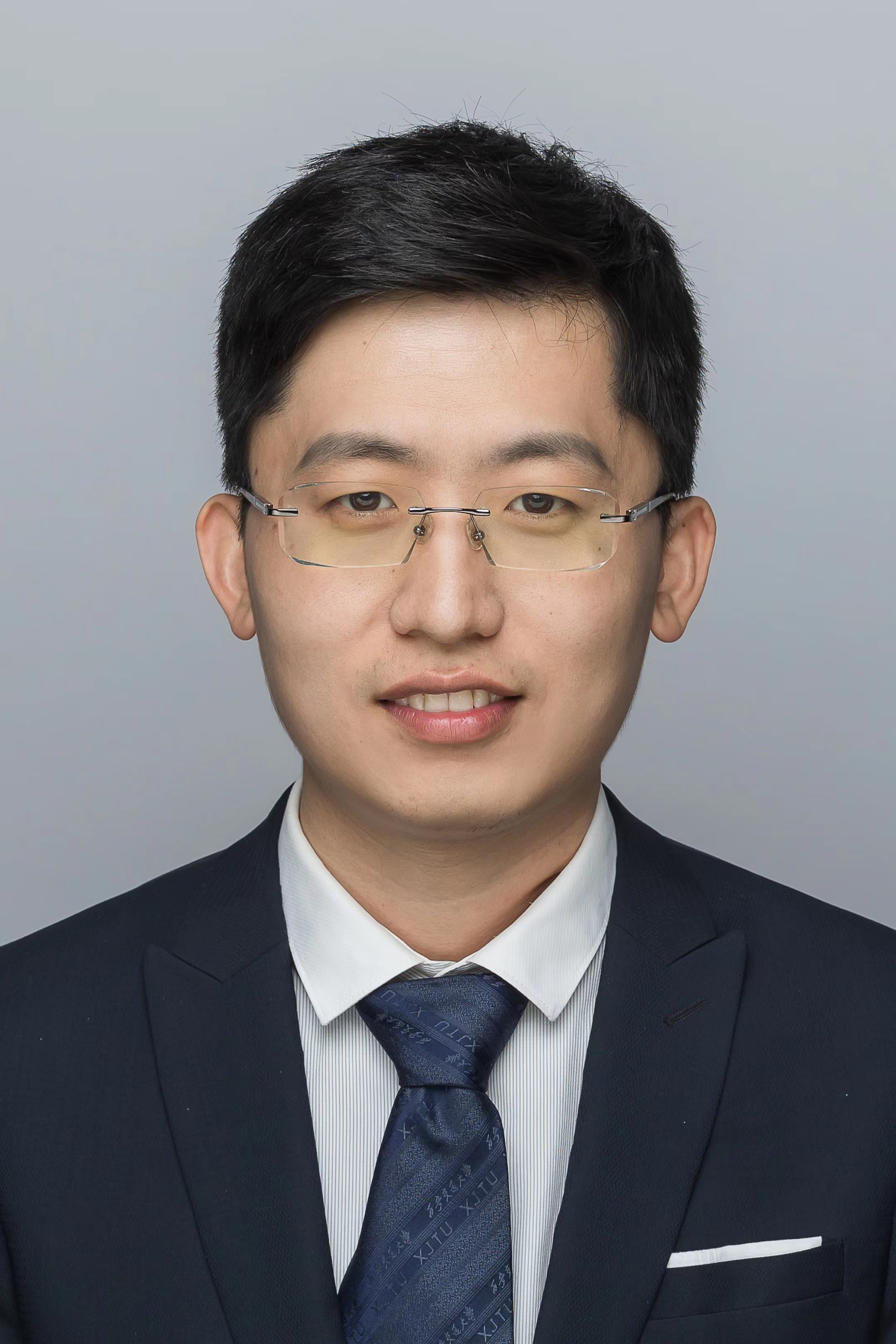}}]{Songhao Yang}
	(S’18-M’19) was born in Shandong, China, in 1989. He received the B.S. and Ph.D. degrees in electrical engineering from the Xi’an Jiaotong University, Xi’an, China, in 2012 and 2019, respectively. Besides, he received the Ph.D. degree in electrical and electronic engineering from Tokushima University, Japan, in 2019. 
	
	Currently, he is an Associate Professor at Xi’an Jiaotong University. His research interest includes power system stability analysis and control.
\end{IEEEbiography}
\begin{IEEEbiography}[{\includegraphics[width=1in,height=1.25in,clip,keepaspectratio]{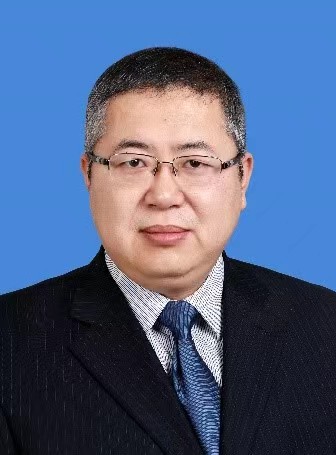}}]{Zhiguo Hao}
	(M’10-SM’23) was born in Ordos, China, in 1976. He received his B.S. and Ph.D. degrees in electrical engineering from Xi'an Jiaotong University, Xi'an, China, in 1998 and 2007, respectively. 
	
	He has been a Professor with the Electrical Engineering Department, Xi'an Jiaotong University since 2018. His research interest includes protection and control of power systems and equipment.
\end{IEEEbiography}

\begin{IEEEbiography}[{\includegraphics[width=1in,height=1.25in,clip,keepaspectratio]{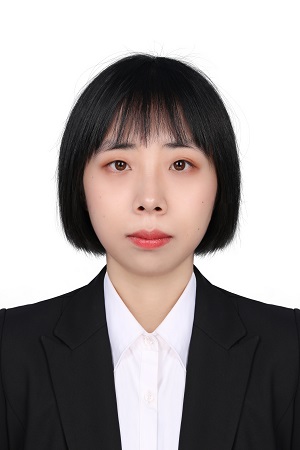}}]{Hongyue Ma}
	(S’23) received the B.S. degree from Harbin University of Science and Technology, Harbin, China in 2015, and M.S. degree from Xi’an Jiaotong University, Xi’an, China in 2020, where she is working toward a Ph.D degree, all in electrical engineering. 
	
	Her research interest includes renewable energy power system analysis and protection.
\end{IEEEbiography}
\begin{IEEEbiography}[{\includegraphics[width=1in,height=1.25in,clip,keepaspectratio]{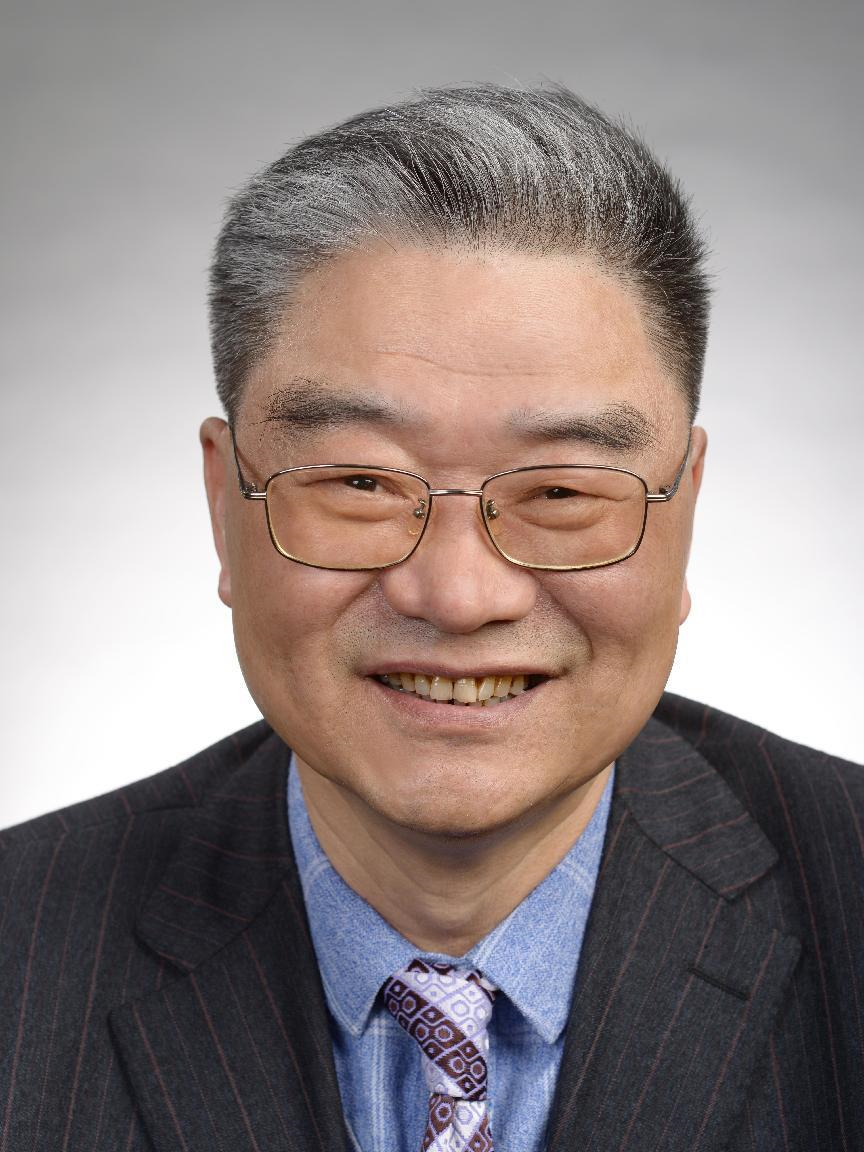}}]{Baohui Zhang}
	(SM’99-F’19) was born in Hebei Province, China, in 1953. He received the M.Eng. and Ph.D. degrees in electrical engineering from Xi’an Jiaotong University, Xi’an, China, in 1982 and 1988, respectively. 
	
	He has been a Professor in the Electrical Engineering Department at Xi’an Jiaotong University since 1992. His research interests are power system analysis, control, communication, and protection.
\end{IEEEbiography}

\end{document}